\documentclass[useAMS,usenatbib]{mn2e}
\usepackage{amssymb}
\usepackage{graphicx}
\DeclareGraphicsExtensions{.jpg,.pdf,.png,.eps,.ps}
\usepackage{url}
\usepackage{color}
\usepackage{amsmath}

\newcommand{\xmmlong}{\textit{XMM-Newton}}

\newcommand{\nh}{\mbox{$N_{\rm H}$}}

\newcommand{\kteff}{\mbox{$kT_{\rm eff}$}}
\newcommand{\tpole}{\mbox{$T_{\rm pole}$}}
\newcommand{\tpinf}{\mbox{$T_{{\rm pole}, \infty}$}}
\newcommand{\rinfty}{\mbox{$R_{\infty}$}}
\newcommand{\rns}{\mbox{$R_{\rm NS}$}}
\newcommand{\mns}{\mbox{$M_{\rm NS}$}}

\newcommand{\nhtt}{\mbox{$N_{H,22}$}}
\newcommand{\chisq}{\mbox{$\chi^2$}}

\newcommand{\Chisq}[3]{$\chi^2_\nu$/dof (prob.) = {#1}/{#2} (#3)}

\newcommand{\Fx}{\mbox{$F_{\rm X}$}}

\newcommand{\xray}{\mbox{X-ray}}

\newcommand{\simlt}{\mathrel{\hbox{\rlap{\hbox{\lower4pt\hbox{$\sim$}}}\hbox{$<$}}}}
\newcommand{\simgt}{\mathrel{\hbox{\rlap{\hbox{\lower4pt\hbox{$\sim$}}}\hbox{$>$}}}}

\newcommand{\approxlt}{\mbox{$\,^{<}\hspace{-0.24cm}_{\sim}\,$}}
\newcommand{\ee}[1]{\mbox{$10^{#1}$}}
\newcommand{\tee}[1]{\mbox{$\times 10^{#1}$}}
\newcommand{\ud}[2]{\mbox{$^{+ #1}_{- #2}$}}

\newcommand{\unit}[1]{\mbox{$\rm\,#1$}}
\def\deg{\hbox{$^\circ$}}

\def\arcsec{\hbox{$^{\prime\prime}$}}
\def\sec{\mbox{$\,{\rm sec}$}}
\newcommand{\G}{\mbox{$\,G$}}
\newcommand{\msun}{\mbox{$\,M_\odot$}}

\newcommand{\km}{\hbox{$\,{\rm km}$}}
\newcommand{\erg}{\mbox{$\,{\rm erg}$}}

\newcommand{\keV}{\mbox{$\,{\rm keV}$}}

\newcommand{\msec}{\mbox{$\,{\rm ms}$}}
\newcommand{\yr}{\mbox{$\,{\rm yr}$}}
\newcommand{\kpc}{\mbox{$\,{\rm kpc}$}}

\newcommand{\persec}{\mbox{$\,{\rm s^{-1}}$}}

\newcommand{\percmsq}{\mbox{$\,{\rm cm^{-2}}$}}

\newcommand{\cgsflux}{\mbox{$\,{\rm erg\,\percmsq\,\persec}$}}
\newcommand{\cgslum}{\mbox{$\,{\rm erg\,\persec}$}}

\title{Modelling of the Surface Emission of the Low-Magnetic Field
  Magnetar SGR~0418$+$5729}

\author[S. Guillot et al.]{S. Guillot$^{1,2}$\thanks{email:
    sguillot@astro.puc.cl}, R. Perna$^{3}$, N. Rea$^{4,5}$,
  D. Vigan\`o$^5$, J. A. Pons$^6$ \\ 
  $^1$ Instituto de Astrof\'{i}sica, Facultad de F\'{i}sica,
  Pontificia Universidad Cat\'{o}lica de Chile, Av. Vicu\~{n}a
  Mackenna 4860, \\782-0436 Macul, Santiago, Chile \\
  $^2$ Department of Physics, McGill University, 3600 rue University
  Montr\'{e}al, QC, Canada H3A-2T8\\
  $^3$ Department of Physics and Astronomy, Stony Brook University,
  Stony Brook, NY 11794-3800, USA \\ 
  $^4$ Anton Pannekoek Institute for Astronomy, University of
  Amsterdam, Science Park 904, Postbus 94249, Amsterdam, The
  Netherlands \\ 
  $^5$ Institute of Space Sciences (CSIC-IEEC), Campus UAB, Carrer de
  Can Magrans s/n, 08193, Barcelona, Spain \\
  $^6$ Departament de Fisica Aplicada, Universitat d'Alacant,
  Ap. Correus 99, E-03080 Alacant, Spain\\}
\begin{document}


\maketitle

\label{firstpage}

\begin{abstract}
  We perform a detailed modelling of the post-outburst surface
  emission of the low magnetic field magnetar SGR~0418+5729.  The
  dipolar magnetic field of this source, $B=6\tee{12}\G$ estimated
  from its spin-down rate, is in the observed range of magnetic fields
  for normal pulsars.  The source is further characterized by a high
  pulse fraction and a single-peak profile.  Using synthetic
  temperature distribution profiles, and fully accounting for the
  general-relativistic effects of light deflection and gravitational
  redshift, we generate synthetic \xray\ spectra and pulse profiles
  that we fit to the observations.  We find that asymmetric and
  symmetric surface temperature distributions can reproduce equally
  well the observed pulse profiles and spectra of SGR~0418.
  Nonetheless, the modelling allows us to place constraints on the
  system geometry (i.e. the angles $\psi$ and $\xi$ that the rotation
  axis makes with the line of sight and the dipolar axis,
  respectively), as well as on the spot size and temperature contrast
  on the neutron star surface.  After performing an analysis iterating
  between the pulse profile and spectra, as done in similar previous
  works, we further employed, for the first time in this context, a
  Markov-Chain Monte-Carlo approach to extract constraints on the
  model parameters from the pulse profiles and spectra,
  simultaneously.  We find that, to reproduce the observed spectrum
  and flux modulation: (a) the angles must be restricted to
  $65\deg\simlt\psi+\xi\simlt125\deg$ or $235\deg\simlt\psi+\xi\simlt
  295\deg$; (b) the temperature contrast between the poles and the
  equator must be at least a factor of $\sim6$, and (c) the size of
  the hottest region ranges between 0.2--0.7\km\ (including
  uncertainties on the source distance).  Last, we interpret our
  findings within the context of internal and external heating models.

  \vspace{0cm}
\end{abstract}



\section{Introduction}
\label{sec:intro}

Isolated neutron stars (NSs) are characterized by a bewildering
variety of astrophysical manifestations. Among those, particularly
intriguing is a class of sources characterized by long periods ($P\sim
2-11\sec$) and high quiescent X-ray luminosities ($L_x\sim
10^{33}-10^{35}\cgslum$), generally larger than their entire reservoir
of rotational energy \citep{mereghetti08}.  These sources,
historically classified as Anomalous X-ray Pulsars (AXPs) and Soft
Gamma-Ray Repeaters (SGRs), often display stochastic bursts of X-rays,
releasing energies $\sim 10^{39}-10^{41}\erg$ in timescales of seconds
or less, and sporadic though very energetic $\gamma$-ray flares, with
typical energetics $\sim 10^{44}-10^{45}\erg$. Furthermore, AXPs and
SGRs also show long-term outbursts, where their X-ray emission
increase up to several orders of magnitudes in days/weeks, and decays
on timescales of years \citep{rea11}.

The most successful model to explain both the high quiescent X-ray
luminosities, as well as the X-ray bursts, giant $\gamma$-ray flares,
and long-term outbursts, is the {\em magnetar} model
\citep{thompson95,thompson96}, according to which SGRs and AXPs are
NSs endowed with large magnetic fields, $B\sim 10^{14}-10^{15}\G$,
resulting for example from an active dynamo at birth
\citep{kouveliotou98}\footnote{An updated compilation of the measured
  dipolar fields can be found here:
  \url{http://www.physics.mcgill.ca/~pulsar/magnetar/main.html};
  \citep{olausen14}.}.  After a magnetar is born, the internal
magnetic field is subject to a continuous evolution through the
processes of Ohmic dissipation, ambipolar diffusion, and Hall drift.
In the crust, magnetic stresses are generally balanced by elastic
stresses.  However, as the internal field evolves, local magnetic
stresses can occasionally become too strong to be balanced by the
elastic strength of the crust, which hence breaks, and the extra
stored magnetic/elastic energy becomes available for powering the
bursts and flares (\citealt{thompson95,thompson96}).  Alternatively,
the ourbursts could be triggered by instabilities in the external
magnetic flux tubes \citep{beloborodov14,link14,lyutikov15}.

While the magnetar model has been very successful in explaining some
general features of the triggering mechanism of bursts and flares, the
discovery in 2010 of an outburst from SGR~0418+5729 \citep[SGR~0418
  hereafter,][]{vanderhorst10,esposito10}, a NS with dipolar magnetic
field $B_{\rm dip}=\left(6\pm2\right)\tee{12}\G$ \citep{rea10,rea13},
lower than those of most magnetars, clearly showed that the overall
picture was not complete.  Following this discovery, two more sources
showing magnetar-like activity, but with ``non-magnetar''-like $B$
fields have then been discovered \citep{rea12,scholz12,rea14,zhou14}.

During the last few years, a number of investigations have been aimed
at understanding the physical reasons for the diverse phenomenology of
magnetars
\citep{pons07,pons09,aguilera08b,aguilera08a,kaspi10,perna11,vigano13}. In
particular, a suite of magnetothermal simulations highlighted the
importance of a hidden toroidal field in determining the observational
manifestations of a NS \citep{pons11,vigano13}.  A NS with a low
dipolar component of the magnetic field (as inferred from timing
measurements) could still display an outbursting behaviour and an
enhanced quiescent X-ray luminosity if endowed with a much stronger
internal toroidal field \citep{turolla11,perna11,rea12}.

However, the presence of a strong toroidal field remains hidden from
timing measurements.  Nonetheless, this component of the field leaves
its strong imprint on the surface temperature of the NS
\citep[e.g.,][]{shabaltas12,perna13,geppert14}, which can be probed by
means of phase-resolved spectral analysis of the quiescent X-ray
emission.  The goal of this work is to perform such an analysis on the
post-ouburst emission of the low-$B$ field source SGR~0418 with the
aim of constraining its surface temperature distribution and, in turn,
gaining some insight into the topology of the magnetic field in the NS
crust and into whether an additional (external) source of heating may
be needed (e.g., see \citealt{bernardini11} for a similar analysis on
the quiescent emission of the transient magnetar XTE~J1810-197).

The structure of this article is as follows: Section~\ref{sec:model}
presents the surface emission model used and details the computation
of the spectra.  Section~\ref{sec:analysis} describes the data
reduction and the spectral and timing analyses performed.  The results
of the modelling are presented in Section~\ref{sec:results}, first
using an iterative analysis, then with a simulateneous analysis of the
pulse profile and spectrum.  In Section~\ref{sec:magneto}, we discuss
our findings within the context of magnethotermal models, and finally,
Section~\ref{sec:ccl} concludes this article with a short discussion
and a summary of the results.

\section{Spectral models for SGR~0418+5729}
\label{sec:model}

\subsection{Family of temperature profiles for SGR~0418+5729}
The timing properties of SGR~0418, as well as its X-ray luminosity,
are consistent with those of an evolved magnetar which experienced
substantial field decay but still retains a strong enough internal
toroidal field. In particular, \cite{turolla11}, using the
magnetothermal code by \cite{pons09}, found that the quiescent
luminosity of the source and its timing properties are compatible with
those of an old NS born with a super-strong magnetic field which
underwent significant decay over a time $\approx 10^6\yr$.  More
recently, following a refined timing solution \citep{rea13}, the
realistic age\footnote{This is to be compared with the charasteristic
  age $\tau_{c}=P/2\dot{P} = 35\unit{Myr}$ of SGR~0418.} of SGR~0418
was estimated to be $\sim550\unit{kyr}$ with the state-of-the-art
magnetothermal evolution model of \cite{vigano13}.  The initial
strength of the dipolar component was estimated to be in the range
$B^{0}_{\rm dip}\sim 1-3\times 10^ {14}\G$ (\citealt{turolla11};
\citealt{rea13}; a larger value would have spun down the pulsar too
much during its estimated lifetime).  However, in order to display a
non-negligible outburst rate, they argued (using the results from the
simulations of \citealt{perna11} and \citealt{pons11}) that the
internal toroidal field at birth had to be much larger.

As discussed in Section~\ref{sec:intro}, independent constraints on
the crustal magnetic topology can be obtained through the imprints of
the magnetic field on the surface temperature of the star.  In the NS
crust, the coupled evolution of the temperature and the magnetic field
gives rise to an anisotropic temperature profile, with the degree of
anisotropy being controlled by the ratio between thermal conductivity
along and across the field lines (for a complete description, see
\citealt{vigano13}).  As the NS ages, the magnetic field in its crust
evolves under the combined influence of the Lorentz force (causing the
Hall drift, see
e.g. \citealt{goldreich92,hollerbach02,hollerbach04,cumming04,pons07,vigano12,gourgouliatos14})
and the Joule effect (responsible for Ohmic dissipation). For a field
at birth which is predominantly poloidal, the symmetry with respect to
the equator is maintained throughout the evolution. However, the
presence of strong internal toroidal components can drastically change
this topology.  If the toroidal field is dipolar, the equatorial
symmetry is broken during the evolution due to the Hall term in the
induction equation \citep[e.g.,][]{vigano13} which leads to a complex
field geometry with asymmetric north and south hemispheres. Hence
asymmetric temperature profiles are produced, with a degree of
anisotropy strongly dependent on the initial toroidal field strength.
On the other hand, a strong internal quadrupolar toroidal field will
maintain the symmetry between the two hemispheres, while increasing
the temperature contrast between the hotter and coler regions on the
NS surface. A suite of magnetothermal simulations by \cite{perna13}
and \cite{geppert14} illustrated the effects of a strong internal
toroidal field as the NS ages. Temperature differences by more than a
factor of two between the two hemispheres could be produced in evolved
objects, though the specific quantitative details depend on some
elements of the microphysics such as the magnitude of the conductivity
and the composition of the envelope. Also the mass of the star
influences the rate of cooling, as well as the extent by which the
field penetrates into the core \citep{vigano13}.

For the purpose of this work, starting with a large set of
magnetothermal simulations by varying all the relevant parameters
described above and trying a one-by-one fit to each of them is
impractical, especially in light of the fact that the magnetothermal
simulations are computationally expensive.  Therefore, we adopt the
following strategy:
\begin{description}
\item[(a)] We use the observed properties of the pulsed profile,
  together with the qualitative predictions for the field
  strength/configuration expected for an evolved magnetar with the
  characteristics of SGR~0418, to produce a family of analytical,
  parameterized temperature profiles, with which spectra and pulsed
  profile are fitted. The fits largely restrict the allowed parameter
  space for the temperature profiles.
\item[(b)] We then discuss our results within the context of
  magnetothermal simulations, as well as possible external heating, to
  identify the most likely physical scenario for the production of
  temperature profiles consistent with those derived from the fits.
\end{description}

\begin{figure}
  \centering
  \makebox[0cm]{\includegraphics[width=8cm]{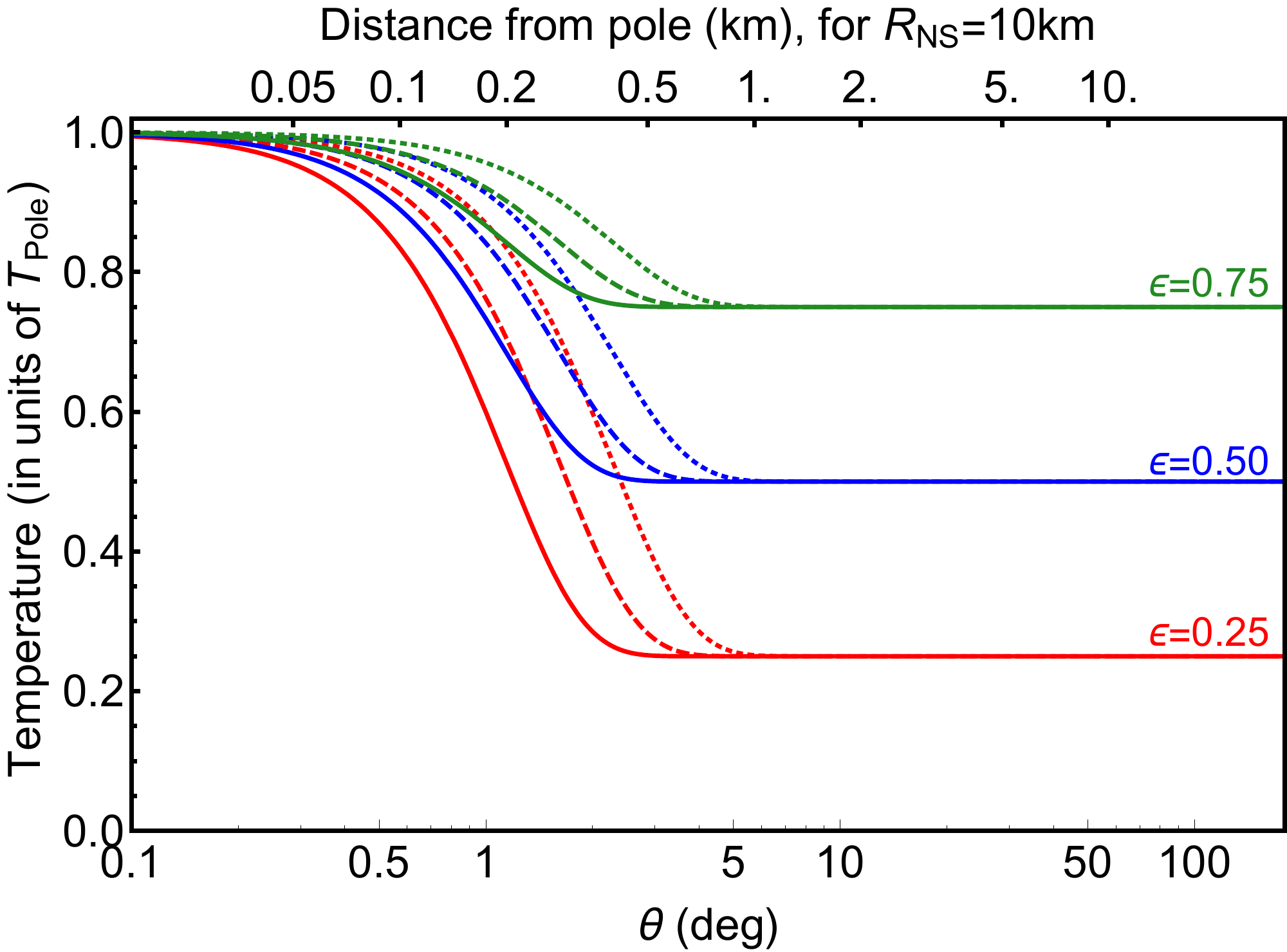}}
  \caption{Synthetic profiles of the temperature (in units of \tpole)
    on the surface of the NS as a function of the angle from the
    magnetic pole ($\theta=0\deg$).  A range of profiles is shown
    here, with different values of $\epsilon$ and $n$ (see
    Equation~\ref{eq:Tprofile}): Red, Blue and Green curves correspond
    to $\epsilon=$ 0.25, 0.50, 0.75, respectively.  The dotted, dashed
    and solid curves correspond to $n =$ 5000, 10000, 20000,
    respectively.}
  \label{fig:TempDistrib1}
\end{figure}

Single-pulse profiles with high pulse-fractions (PFs) such as that
resulting from the observations of SGR~0418 cannot be readily
explained by pairs of antipodal hot spots on the surface of the
rotating NS. In fact, \citet{beloborodov02} showed that two
isotropically emitting, infinitesimal hot spots on the surface of a NS
(of 1.4 solar masses and 10~km in radius) cannot produce PFs larger
than about 12\%. \citet{poutanen06} generalized the calculation to
allow for anisotropy in the local emission, and obtained PFs of at
most a few tens of percent. Similarly, \citet{dedeo01} studied the
modulation level of the emitted flux as a function of the hot spot
size, for two antipodal spots. They found that, for a NS of $1.4$
solar masses, radius 12.3~km, and a temperature contrast of $9$
between the spots and the rest of the surface, the PF can reach a
maximum of $\sim 55$\% if the local emission is strongly
pencil-beamed, with an intensity $\propto \cos^3\delta$, where
$\delta$ is the angle of the emitted photon with the normal to the
surface. However, a more recent study was performed by
\citet{shabaltas12}, using a family of parametrized temperature
profiles with symmetry with respect to the equator. By employing a new
method to efficiently compute the radiation intensity from different
patches of a NS surface with arbitrary magnetic fields and effective
temperatures, they were able to produce synthetic X-ray light curves
for a variety of geometric configurations. In contrast to previous
works, they were able to generate single-peaked profiles with PFs as
high as $\sim 60\%$. They interpreted their findings as the result of
diffuse hot spots of finite size (with varying temperatures), combined
with the beaming due to anisotropic photon opacities in magnetic
fields. We note that, consistently with previous work, their single
pulsed profiles from symmetric temperature distributions displayed a
plateau around the flux minimum. Using the same methodology of
\citet{shabaltas12}, \citet{storch14} modelled the high X-ray PFs of
the pulsar PSR B0943$+$10 by means of two asymmetric hot spots, a
smaller one with stronger $B$ field, and a larger one with smaller $B$
field. Once again, the beaming effect of the atmosphere was found to
be a crucial ingredient in producing a high modulation level.

We note that both the source Kes~79 analyzed by \citet{shabaltas12},
as well as the pulsar PSR~B0943$+$10 modelled by \citet{storch14}, had
relatively well measured distances, and in addition, PSR~B0943$+$10
had the viewing geometry constrained by radio observations. As a
result, the area of the thermally emitting region could be well
determined. On the other hand, the distance to SGR~0418 is rather
uncertain. If located in the Perseus arm (which is in its direction),
then the distance would be about
2\kpc\ \citep{vanderhorst10}. However, there is no independent
observational link to such an association.  Because of this
uncertainty, we choose to leave the distance to the source as a free
parameter, which, as discussed in the following (see
Section~\ref{sec:iterative}), will turn out to be correlated with the
temperature profile on the NS surface, and especially with the size of
the spot.  In addition, given the characteristics of the source, and
in particular its symmetric pulse profile (i.e. no evidence for a
plateau around the minimum), as well as the very high PFs ($\sim
80\%$, and even consistent with $\sim 100\%$ above 1.2~keV), we focus
our exploration on a family of temperature profiles asymmetric with
respect to the equator, with one hemisphere hotter than the
other. This choice is less restrictive in that it allows a wider range
of sizes for the hotter component (and correspondingly a wider range
of distances). However, we will also explore whether, similarly to the
cases studied by \citet{shabaltas12} and \citet{storch14}, the
viewing/emission geometry compatible with the data allows for the
presence of a second antipodal hot region.

We begin by exploring the following family of temperature profiles
\begin{equation}
T\left(\theta\right) = \epsilon \; \tpole + \left( 1-\epsilon
\right)\;\tpole \; \cos^n \left( \frac{\theta}{2}
\right),
\label{eq:Tprofile}
\end{equation}
where $\theta$ is the azimuthal angle, $\epsilon$ represents the
fractional temperature difference between the two poles, and $n$ is an
integer power which measures the gradient of the temperature decline
between the two poles.  Some representative profiles from this family
are displayed in Figure~\ref{fig:TempDistrib1}, for different values
of the parameters $\epsilon$ and $n$.

Similarly, the symmetric profiles (i.e., a pair of antipodal
  spots) are parametrized as
 \begin{equation}
T\left(\theta\right) = \epsilon \;  \tpole + \left( 1-\epsilon
\right)\;\tpole \; \cos^n \left({\theta}\right),
\label{eq:Tprofile2}
\end{equation}
with $n$ constrained to be even.

We note that there is a certain degree of degeneracy between the
gradient of the temperature profile, the local emission pattern of the
radiation, the compactness $\mns/\rns$ of the NS, and the viewing
geometry. For the local emission, we assume blackbody radiation with
an analytical prescription for its beaming. We realize that this is an
approximation in the case that the star has an atmosphere, which
processes the surface radiation creating a distorted blackbody and
modifying the emission pattern. The only correct way to approach this
problem would be to generate a magnetized atmosphere model for each
patch of the star (different $B$ and $T$). To the best of our
knowledge, this has been done only with an approximate method to
produce realistic pulsed profiles coupled with analytical
temperature/magnetic profiles \citep{shabaltas12,storch14}. However,
it is impractical to implement this method while formally fitting data
since it would require the computation of a much larger number of
atmospheric profiles than done in those papers (we will be minimizing
over a wide grid of parameters for both the temperature profile and
the viewing geometry).  On the other hand, using a single model
atmosphere (i.e., a single strength for $B$ and one direction,
perpendicular to the normal) on the entire surface -- which is
reasonably easy to do in a fit -- would be generally incorrect, and
especially so if the object has a significant non-radial component of
the external magnetic field.  Therefore, given that the detailed
magnetic topology of SGR~0418 is not apriori known, and the fact that,
even with a perfect spectral computation there would still remain a
degree of degeneracy in the pulsed profiles with the compactness ratio
of the star (which we assume fixed at some typical value), and the
viewing geometry, which we constrain by fitting the pulse profile, we
adopt the simplest approach of assuming local blackbody emission with
a parametrized form for the beaming which captures the
'limb-darkening' effect of magnetized, light-element atmospheres (see
also \citealt{bogdanov14} for a similar approach). However, in order
to quantify the dependence of the results we obtain on the presence of
beaming, we will also repeat part of the analysis for isotropic
emission.

\subsection{General-relativistic, phase-dependent spectra}

The general relativistic, phase-dependent spectra are calculated using
the formalism developed by (\citealt{page95}, see also
\citealt{pechenick83, pavlov00a}).  The intense gravitational field of
the star bends the photon trajectories: a photon emitted at an angle
$\delta$ with the normal to the NS surface will reach a distant
observer if generated at an angle $\theta_v$ with respect to the
viewing axis, where
\begin{eqnarray}
\theta_v(\delta) = \int_0^{\frac{R_s}{2R}}
x\left[\left(1-\frac{R_s}{R}\right)\left(\frac{R_s}{2R}\right)^2-(1-2u)u^2
  x^2\right]^{-\frac{1}{2}}\mbox{d}u\,.
\label{eq:teta}
\end{eqnarray}

\noindent
In the above equation, $x\equiv\sin\delta$, $R_{s}\equiv 2GM/c^2$ is
the Schwarzschild radius of the star, and $R$ and $M$ are,
respectively, its radius and mass.  { Here we adopt a canonical mass
  $M=1.4\msun$ and radius \rns=10\km.}  These values yield a
gravitational redshift $(1-R_{s}/R)^{1/2}=0.766$.  Correspondingly, a
distant observer would measure a radius of $\rinfty =
R\times({1-{{R_s}/{R}}})^{-1/2}=13.1\km$, and a surface temperature of
$T_\infty=T\times (1-R_{s}/R)^{1/2}$.  In the following, we will quote
fit results for the pole temperature in terms of the redshifted
values, \tpinf.

Let $\Omega(t)$ be the modulus of the NS angular velocity, and let us
define the phase angle $\gamma(t)=\int \Omega(t) dt$ as the azimuthal
angle subtended by a reference vector ${\bf \hat{n}}$ around the axis
of rotation. As the reference vector, we choose the polar axis, in the
coordinate system in which $T(\theta=0)\equiv \tpole$.  For a dipolar
component of the magnetic field, this would correspond to the magnetic
axis; for the toroidal component of the field, it represents the
symmetry axis.

As the star rotates, the angle between the vector ${\bf \hat{n}}$ and
the line of sight is given by
\begin{eqnarray}
\label{eq:alpha}
\alpha(t) =
\arccos\left[\cos\psi\cos\xi+\sin\psi\sin\xi\cos\gamma(t)\right]\,,
\end{eqnarray}
where we indicated by $\psi$ and $\xi$ the angles that the rotation
axis respectively makes with the line of sight and ${\bf\hat{n}}$ (see
Figure 1 in \citealt{perna08} for a graphical representation of the
viewing geometry). Note that there is a degeneracy between the two
angles $\psi$ and $\xi$, i.e., they can be exchanged without altering the
pulse-profile or the spectra.

The phase-dependent spectrum measured by the observer as the star
rotates is obtained by integrating the local emission over the entire
observable surface. Accounting for the effect of gravitational
redshift of the emitted radiation, this integral takes the form
\begin{eqnarray}
F(E_\infty,\alpha)=\frac{E_\infty^2}{c^2h^3}\frac{R_\infty^2}{D^2}\;
\int_0^1 2x \int_0^{2\pi}
I\{T[\theta_v(x),\phi_v],E\}\,\mbox{d}\phi_v\mbox{d}x\,,
\label{eq:flux}
\end{eqnarray}
where $D$ is the distance, $E_\infty=E({1-{{R_s}/{R}}})^{1/2}$ is the
energy measured by a distant observer, the spectral function
$I[T(\theta_v, \phi_v), E]$ describes the dimensionless distribution
of the locally emitted photons (the blackbody function here), and
$(\theta_v,\phi_v)$ are the coordinates on the surface relative to the
line of sight.

As discussed above, to emulate the effect of a magnetized,
light-element atmosphere, we assume the local photon intensity to
emerge in a pencil-beaming pattern with respect to the normal to the
surface, which we model as $f(\delta)\propto\cos^p\delta$, with a
beaming intensity $p=1$ as a closer match to that of realistic,
magnetized atmosphere models \citep{vanadelsberg06}\footnote{These
  authors particularly discussed the important effect of vacuum
  polarization on the emergent radiation pattern. They showed how,
  without the inclusion of this effect, the emitted radiation exhibits
  a characteristic beaming pattern, with a thin ‘pencil’ shape at low
  emission angles and a broad ‘fan’ at large emission
  angles. Inclusion of vacuum polarization tends to reduce the gap and
  lead to a featureless, broad pencil beaming pattern (especially so
  for stronger fields).}.

From Equation~(\ref{eq:flux}), other useful quantities for comparison
to observations can then be readily computed. In particular, the
phase-averaged spectrum is given by
\begin{equation}
F_{\rm ave}(E_\infty) = \frac{1}{2\pi}\;\int_0^{2\pi} \;F[E_\infty,
  \alpha(\gamma)]\; d\gamma\;,
\label{eq:fave}
\end{equation}
while the pulse profile in a given (observed) energy band,
$\{E_{1,\infty},E_{2,\infty}\}$, is
\begin{equation}
F(\gamma) = \int_{E_{1,\infty}}^{E_{2,\infty}}\; F[E_\infty,
  \alpha(\gamma)] \; dE_{\infty}.
\label{eq:fband}
\end{equation}
The PF is defined as
\begin{equation}
{\rm PF} = \frac{F_{\rm max}(\gamma)- F_{\rm min}(\gamma)}{F_{\rm
    max}(\gamma)+ F_{\rm min}(\gamma)}\;,
\label{eq:pf}
\end{equation}
where the phases corresponding to the maximum and minimum of the flux
will generally vary depending on the temperature distribution on the
NS surface.

\begin{table}
 \centering
 \begin{minipage}{140mm}
   \caption{\xmmlong\ Observations of SGR~0418$+$5729}
   \begin{tabular}{ccc}
     \hline
     ObsID & Start     & Usable time\\
           & Time (TT) & (ksec)\\
     \hline
     0693100101 & 2012-08-25 14:18:08 & 58.7 \\
     0723810101 & 2013-08-15 18:06:25 & 32.5 \\
     0723810201 & 2013-08-17 20:57:17 & 36.0 \\
     \hline

     \label{tab:ObsID}
   \end{tabular}
 \end{minipage}
\end{table}

\section{Data reduction and description of analyses}
\label{sec:analysis}

\subsection{\xmmlong\ observations}

We use in this work two recent observations of SGR~0418 with \xmmlong,
acquired in August 2012 and August 2013 (see Table~\ref{tab:ObsID}),
when the source appears to have approached quiescence with a stable
flux.  All observations were acquired with the EPIC-pn camera
\citep{struder01} in Large Window mode, i.e., with a time frame of
48\msec, and with the EPIC-MOS cameras \citep{turner01} in Small
Window mode, with a time resolution of 300\msec.  For the timing
analysis, data from both pn and MOS cameras are used, while only the
pn spectrum is used for the spectral analysis\footnote{The
  uncertainties due to the cross-calibrations of the pn and MOS
  effective areas compensate for the gain in signal-to-noise ratio
  when adding the MOS spectra (see \citealt{read14}).}.

Standard reduction procedures are applied, using {\tt epchain} in the
\emph{XMMSAS} v13.5 package, together with the latest calibration
files.  Photon events times of arrival are corrected to the Solar
System barycenter, using {\tt barycen}, before the phases of all
events are calculated using the best-fit \xray\ timing solution
reported in \cite{rea13}: $P=9.07838822\sec$,
$\dot{P}=4\tee{-15}\sec\persec$ on MJD 54993.0.  The data were checked
for background flares, which were removed to limit contamination.  The
resulting usable exposure times for each observation are listed in
Table~\ref{tab:ObsID}.  Finally, events are filtered in the
0.3--10.0\keV\ range with the PATTERN $\leq$4 and FLAG = 0
restrictions.  The observed flux of both observations is consistent
with being constant, confirming that SGR~0418's luminosity and surface
temperatures have not significantly changed between observations.

\begin{figure}
  \centering
  \makebox[0cm]{\includegraphics[width=8cm]{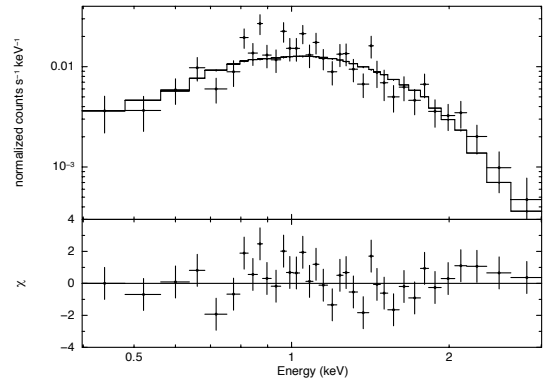}}
  \caption{Phase averaged spectral fit with the model described in
    Section~\ref{sec:model} (see Equation~\ref{eq:flux}), with
    $n=12783$, $\epsilon=0.08$ and $(\psi,\,\xi)=(13\deg,84\deg)$, the
    most likely values from Section~\ref{sec:mcmc}.  The spectral
    shape is mostly driven by the pole temperature \tpole\ and the
    absorption column density $N_{\rm H}$, while the flux depends on
    $n$ and $\epsilon$ with weaker dependence on $\psi$ and $\xi$.
    The fit is statistically acceptable with \Chisq{1.28}{34}{0.13}.}
  \label{fig:spec1}
\end{figure}

\subsection{Spectral analysis}

Phase-averaged spectra are extracted from the two observations in the
0.3--10\keV\ range using 20\arcsec\ circular regions. Background
spectra are extracted from 90\arcsec\ circular regions devoid of
\xray\ sources.  The response matrices were generated using the tools
{\tt rmfgen} and {\tt arfgen}.  The extracted pn spectra are then
combined into a single spectra using {\tt epicspeccombine},
after verification that the flux was consistent with being constant.
Finally, events are binned for the purpose of the phase-averaged
spectral analysis with a minimum of 20 counts per bins.  For the
spectral analysis performed in {\tt XSPEC v.12.8} \citep{arnaud96},
3\% systematics are added in each spectral bin to account for
uncertainties in the absolute flux calibration of the instrument
\citep{guainizzi14}.  The spectral model used, described in
Section~\ref{sec:model}, is modulated by \xray\ absorption using the
{\tt wabs} model \citep{morrison83}.  The overall normalization factor
of the fit, convoluted with the uncertain distance to the source, is
left free to vary.  All errors from the spectral analysis are at the
90\% confidence level.  The spectrum and best-fit model are shown in
Figure~\ref{fig:spec1}, and the results are presented in
Section~\ref{sec:results}.

\subsection{Pulse Profile Analysis}

The phases resulting from the folding at the timing solution given
above are used to generate the pulse profile.  Furthermore, because of
the low count rates in the observations, we limit the analysis to 10
phase bins.  The errors in the number of counts in each bin are
Poisson errors.  The average background number of counts is subtracted
in each phase bin.  For the background subtraction, the background
regions used have the same size as the SGR~0418 source regions, and
are devoid of any other detected \xray\ source.

The PF is calculated according to Equation~(\ref{eq:pf}), where
$F_{\rm max}(\gamma)$ and $F_{\rm min}(\gamma)$ are the maximum and
minimum \xray\ fluxes measured in the phase bins.  The PF for the
cumulative 0.3--10\keV\ band is $PF = 0.78\pm0.09$.  While first using
the single 0.3--10\keV\ band for the computation of the pulsed profile
and modelling, we then also considered splitting the full energy range
into smaller bands, since this carries a higher constraining power for
the underlying model. However, we found that splitting in more than
two bands would result in a too low count rate per band. Hence we
limited the split to only two bands, 0.3--1.2\keV\ and 1.2--10.0\keV,
which were chosen to have approximately similar count rates.  We found
the PF in the lower energy band to be $PF_{0.3-1.2\,{\rm keV}} =
0.62\pm0.10$, while the higher band $PF_{1.2-10.0\,{\rm keV}}$ was
consistent with $1.0$.  As expected, the PF in the high-energy band is
higher than in the low-energy band.

\section{Constraining the surface temperature through spectral and pulse profile modelling}
\label{sec:results}

\subsection{Iterative fitting of spectra and pulse profiles}
\label{sec:iterative}
Constraining the temperature profile requires a coupled spectral and
timing analysis. The pulsed profiles are most sensitive to the run of
temperature with angle on the NS surface, while the phase-averaged
spectra are most sensitive to the overall flux normalization
(reflected in \tpole\ and on the size of the spot, for a fixed NS
radius), and to the amount of absorption, quantified by the column
density of hydrogen \nh, noted \nhtt\ hereafter when expressed in
units of $\ee{22}\unit{{\rm atoms}\percmsq}$. Once \tpole\ and
\nh\ are measured from the spectra, fitting the synthetic pulse
profiles (computed via Equation~\ref{eq:fband}) to the observed pulse
profiles allows to constrain the system geometry and the temperature
profile.  Here, we first perform the spectral and timing analysis
iteratively rather than simultaneously.  We will show that this is a
rather good approximation indeed (see also Bernardini et al. 2011).
However, Section~\ref{sec:mcmc} will present and validate a
simultaneous analysis using a Markov-Chain Monte-Carlo approach.

\begin{figure}
  \centering
  \makebox[0cm]{\includegraphics[width=8cm]{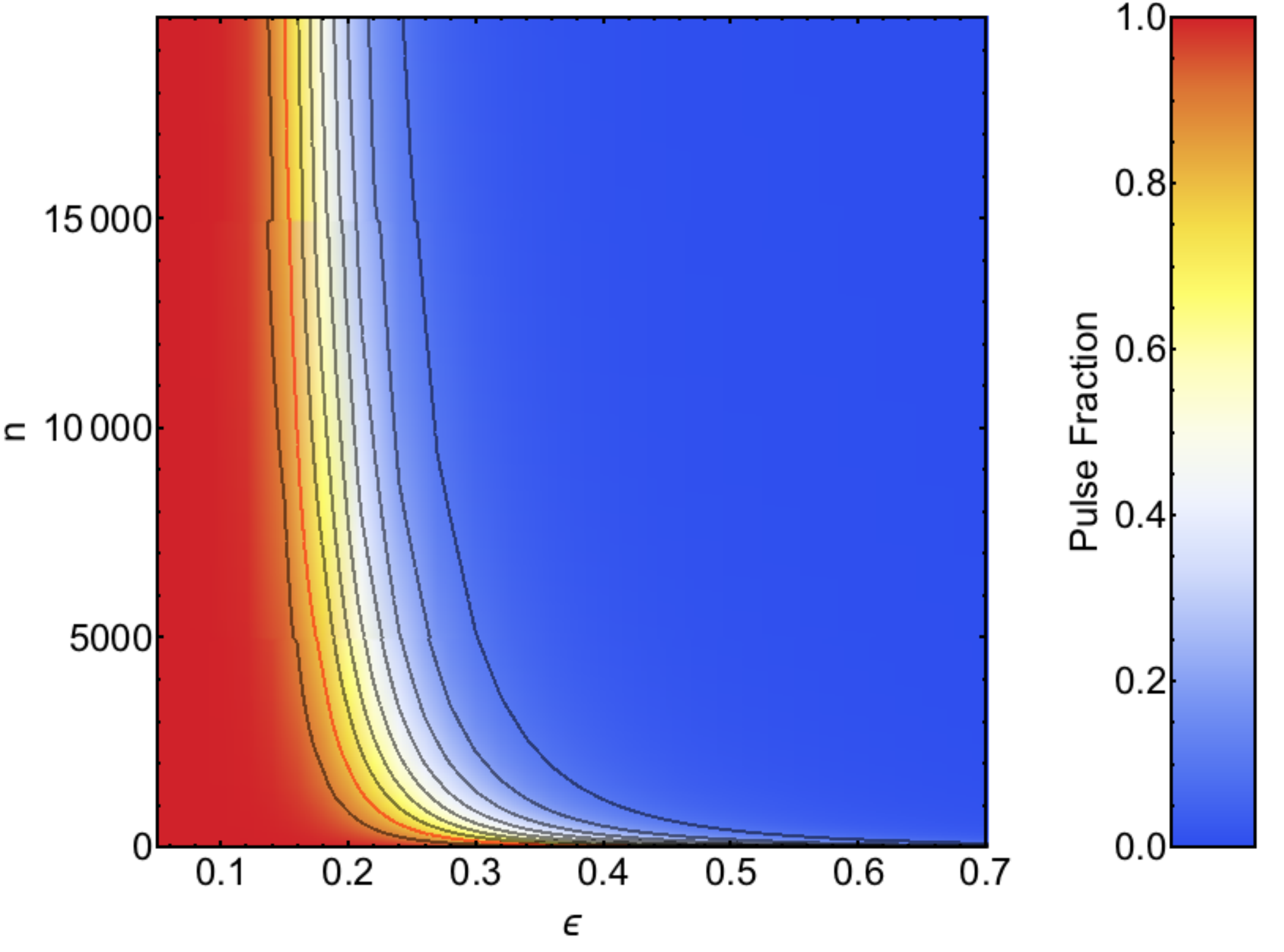}}
  \caption{Pulsed fractions (PFs) obtained as a function of the
    parameters $n$ and $\epsilon$, with the system geometry
    $(\psi,\,\xi)=(90\deg,90\deg)$, which maximizes the PF.  Contour
    lines indicate PFs ranging from 0.1 to 0.9, with 0.1 increments;
    the red line indicates the 0.8 PF contour.  Note that while the
    low range of $n$ values can produce PF such as those observed for
    SGR~0418, it fails to accomodate reasonable distances of this
    source (see discussion in Section~\ref{sec:iterative}).}
  \label{fig:TempDistrib3}
\end{figure}

Using the spectral model described in Section~\ref{sec:model},
together with the family of temperature profiles defined by
Eq.~(\ref{eq:Tprofile}), we first obtain an estimate of \tpole\ and
\nh\ from a small selection of surface temperature distributions, for
a viewing geometry $\left(\psi,\xi\right)=\left(90\deg,90\deg\right)$.
In the first step of the iteration the viewing geometry is not
constrained yet, and the choice of an orthogonal rotator is made as a
starting point given the high PF of the source.  We find that for a
fixed distance (e.g.  $2\kpc$), the spectra strictly constrain the
values of $\epsilon$ and $n$ such that the model produces the flux
observed for this source.  This is because these parameters relate to
the fraction of the NS surface which dominates the emission (see
Figure~\ref{fig:TempDistrib1}), and hence to the fit normalization.
In other words, the shape of the observed spectra constrains the
average emission temperature and absorption, while the flux
normalization restricts the size of the spot and the temperature
constrast at the surface.  We note, however, that the distance is
estimated solely based on the association of the magnetar with the
Perseus arm of the Galaxy.  This assumption stems from the position of
SGR~0418 in the direction of the Perseus arm.  Should this association
be fortuitous, the distance to SGR~0418 could be mis-estimated to
lower or higher values.

Such uncertainties in the distance translate into less constrained
parameters $\epsilon$ and $n$.  As an example, a spectral fit with
$\epsilon=0.1$, $n=10,000$ (i.e., a spot size of $0.275\km$,
full-width at half-maximum, FHWM, on a 10\km\ NS) yields a distance of
$2.13\kpc$, while a fit with $\epsilon=0.15$, $n=5,000$ (spot size of
$0.425\km$ FWHM) yields $4.9\kpc$.  However, the parameters
\tpole\ and \nh\ appear relatively stable against change in $\epsilon$
and $n$, or in the viewing geometry.  We find $\tpinf\sim0.32\keV$ and
$\nhtt\sim0.25$ for a wide range of $\epsilon$ and $n$ that lead to
reasonable values of the distance ($d\sim 0.5-4\kpc$), i.e.,
$\epsilon\approxlt 0.10-0.15$ and $n\sim 5000-20000$.  Therefore we
already have a good starting point in this iterative process to
constrain the parameters of the system.  Higher values of $\epsilon$
lead to statistically unacceptable fits.

We note the existence of a secondary \chisq\ minimum, leading to
acceptable fit statistics, in which $\nhtt\sim1.2$ corresponds to
$\epsilon\sim 0.4$, for the same pole temperature.  This could be
explained by the fact that the larger averaged flux created by the
less pronounced contrast (larger $\epsilon$) needs to be heavily
absorbed to fit the data.  However, we reject such a large value of
the absorption, on account of the fitted values of \nh\ from the high
signal-to-noise observations obtained during the outbursts
\citep{rea13}.

With the estimated values of pole temperature and absorption, we can
then explore the $\epsilon$--$n$ parameter space for pairs capable of
accomodating a PF of $\sim0.80$.  Figure~\ref{fig:TempDistrib3} shows
the PF as a function of $\epsilon$ and $n$ for
$\left(\psi,\xi\right)=\left(90\deg,90\deg\right)$.  With such angles,
a $PF\sim0.80$ can be reproduced for a wide range of $n$, while
$\epsilon$ remains essentially constant at $\sim0.15$.

We note that the high measured PF of the source is partly due to the
effect of \xray\ absorption.  More specifically, absorption can
artificially increase the intrisic (unabsorbed) PF since the
low-temperature portion of the NS surface (emitting lower energy, less
pulsed \xray\ photons) is more affected by absorption than the hotter
parts.  The artifical increase in PF due to absorption depends both on
the geometry $\left(\psi,\xi\right)$ and the temperature gradient on
the star\footnote{See \citet{perna00} for an extensive discussion of
  the effect of absorption on the PFs.}.  The emission model of this
analysis can be used to estimate the intrinsic PF of SGR~0418, i.e.,
unaffected by absorption.  Specifically, for $n=10000$ and
$\epsilon=0.15$, the intrinsic PF is 0.36, while it is 0.88 for
$\nhtt=0.25$, hence quite sensitive to \nh.

\begin{figure}
  \centering
  \makebox[0cm]{\includegraphics[width=8cm]{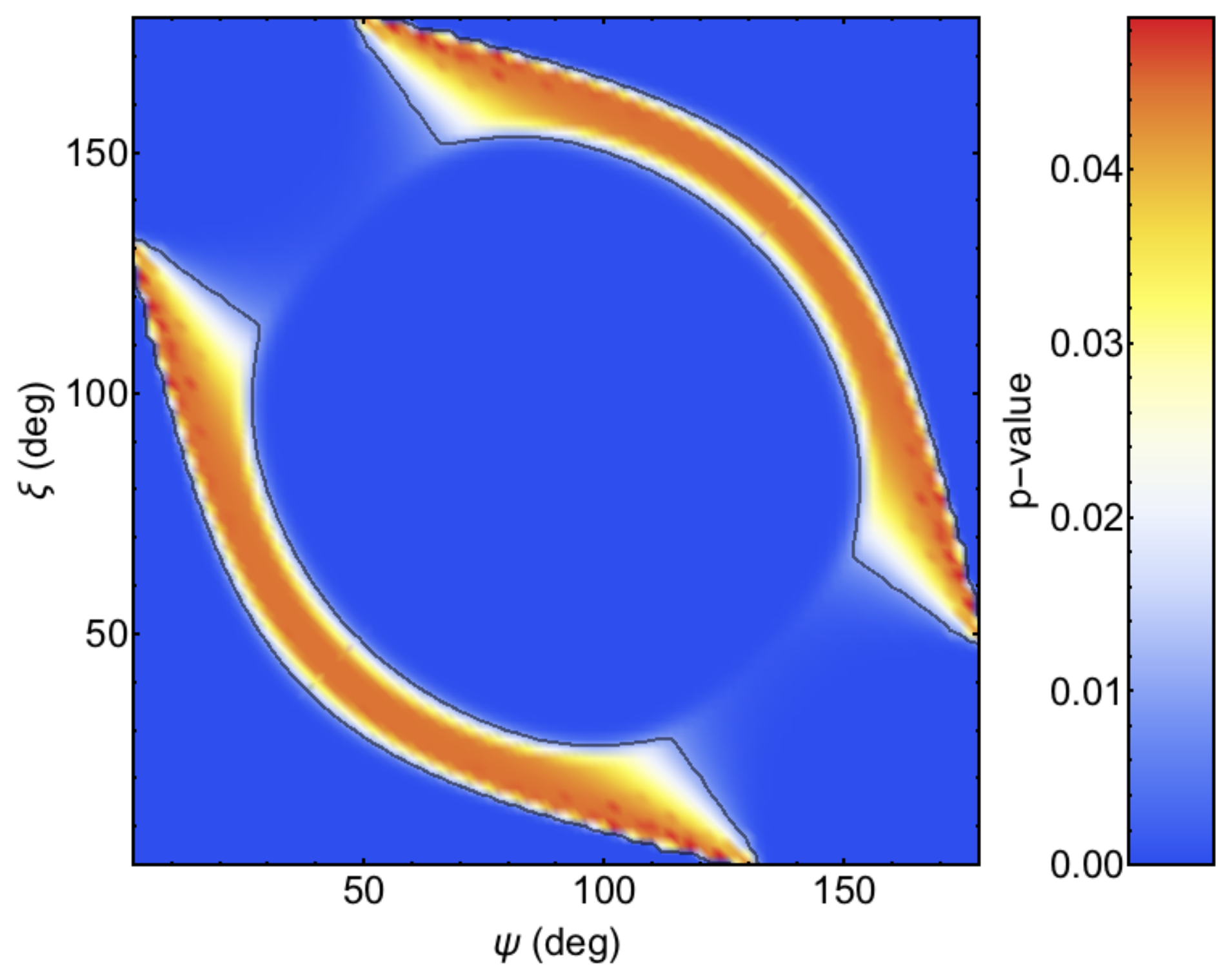}}
  \caption{Projected 2D map of maximum p-values when the two angle
    $\psi$ and $\xi$ vary, in the case of beamed local emission
    (pencil beam $\propto\cos\delta$).  The solid line represents the
    probability $p=0.01$.  The best-fit over the whole 4D-space
    corresponds to a p-value $p=0.05$.}

  \label{fig:beaming}
\end{figure}

Next, choosing $\psi=90\deg$ and $\xi=90\deg$ with the temperature
distribution profile defined by $\epsilon=0.15$ and $n=10000$, we
perform a detailed spectral analysis\footnote{For comparison, we
  provide the effective temperature obtained when fitting a {\tt
    bbodyrad} model: $\kteff_{,\infty} = 0.32\ud{0.05}{0.04}\keV$ and
  $\nhtt=0.18\ud{0.09}{0.08}$ for \Chisq{1.31}{34}{0.10}. Adding a
  {\tt powerlaw} spectral component slightly improves the fit (F-test
  probability $\sim3\%$).}  in order to further refine the value of
the pole temperature \tpole\ and the \xray\ absorption \nh.  An
acceptable fit to the data is obtained with $\tpinf =
0.33\ud{0.05}{0.05}\keV$ and $\nhtt=0.36\ud{0.09}{0.07}$ (see
Figure~\ref{fig:spec1}).  However, a better fit is obtained with
$\epsilon=0.10$ and $n=10000$ leading to $\tpinf =
0.38\ud{0.05}{0.05}\keV$ and $\nhtt=0.22\ud{0.09}{0.07}$ for
\Chisq{1.28}{34}{0.13}.  The flux of SGR~0418 is
$\Fx=2.0\ud{0.4}{0.4}\tee{-14}\cgsflux$ in the 0.3--3.0\keV\ range.
Although these values of $n$ and $\epsilon$ slightly overpredict the
PF (see Figure~\ref{fig:TempDistrib3}), the best-fit \tpole\ and
$N_{\rm H}$ are consistent with those estimated for different values
of $\epsilon$ and $n$ in the first step of this iterative process.

\begin{figure}
  \centering
  \makebox[0cm]{\includegraphics[width=7cm]{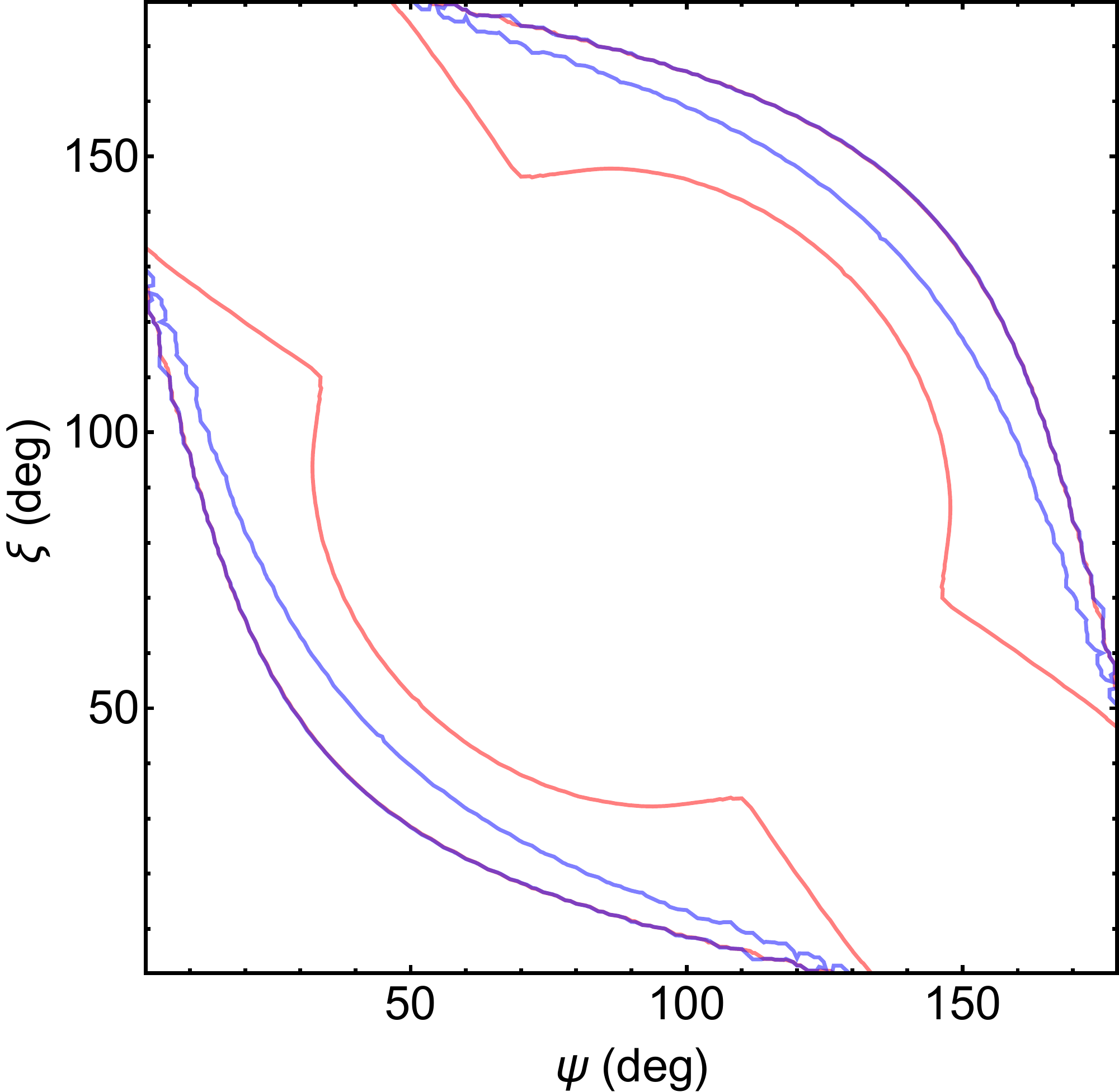}}
  \caption{Same as in Figure~\ref{fig:beaming}, but for the two bands:
    0.3--1.2\keV\ (red) and 1.2--10.0\keV\ (blue).  Only the $p=0.01$
    contours are plotted to allow comparison between the two bands.}
  \label{fig:twobands}
\end{figure}

With the values of $\tpinf=0.38\keV$ and $\nhtt=0.22$ obtained from
the spectral analysis, we next proceed to constrain the system
geometry and the temperature distribution that best reproduces the
pulsed emission from the surface of SGR~0418.  For the analysis that
we describe in the following, we first consider the pulsed profile in
the full energy band 0.3--10\keV.  Using the model described in
Section~\ref{sec:model}, a grid of pulse profiles is generated for
multiple system geometries (varying the angles $\psi$ and $\xi$
between 2\deg and 180\deg, in steps of $2\deg$), and multiple
temperature distribution profiles (by varying $n$ between 5000 and
20000 in steps of 500, and $\epsilon$ between 0.05 and 0.15 in steps
of 0.01).  Each of the synthetic pulse profiles generated for each set
of 4 parameters ($\psi$, $\xi$, $n$, $\epsilon$) is then fitted to the
observed \xmmlong\ pulse profile.  From the \chisq\ values obtained
from these fits, projected 2-dimensional maps of maximum p-values are
obtained\footnote{The p-value \citep[see e.g.,]{vogel14,tendulkar15},
  also called the null hypothesis probability in \emph{XSPEC}, is the
  probability of finding by chance a \chisq\ as large or larger than
  the best-fit \chisq\ under the assumption that the null hypothesis
  is true (i.e., that the model does not describe the data).  The
  p-value is calculated by integrating the \chisq\ probability density
  function with the number of degrees of freedom in the data set
  between the best-fit \chisq\ and $\infty$.}.

Figure~\ref{fig:beaming} shows the maximum p-value maps for the pair
of angles $\psi$-$\xi$.  The fit to the pulsed profile allowed us to
mainly constrain the viewing/inclination geometry.  The parameters $n$
and $\epsilon$ were not constrained by the pulse profile fitting any
further than what was imposed by the spectral analysis (via the
distance requirement).  In other words, $n$ and $\epsilon$ were
essentially unconstrained by the pulse profile fitting in the range
5000--20000 and 0.05--0.15, respectively.

The best-fit resulting from the grid search in the parameter space is
obtained for the following parameters: $\epsilon=0.10$, $n=7500$ for
$\left(\psi,\xi\right)=
\left(\xi,\psi\right)=\left(124\deg,4\deg\right)$, with
\Chisq{1.9}{8}{0.05}.  However, as can be observed in
Figure~\ref{fig:beaming}, a wide range of angles is permitted by the
pulse profile fitting ($p>0.01$), with very different sets of
parameters $n$ and $\epsilon$. Figure~\ref{fig:beaming} also shows
that the angles $\psi$ and $\xi$ are constrained to two narrow bands.
Note the observed symmetry emerges from the symmetry against exchanges
between $\psi$ and $\xi$ (see Equation~\ref{eq:Tprofile}).  The most
conservative constraints on the angles can be summarized by the
intervals: $73\deg\simlt\psi+\xi\simlt140\deg$ and
$220\deg\simlt\psi+\xi\simlt 287\deg$.

\begin{figure}
  \centering
  \makebox[0cm]{\includegraphics[width=8cm]{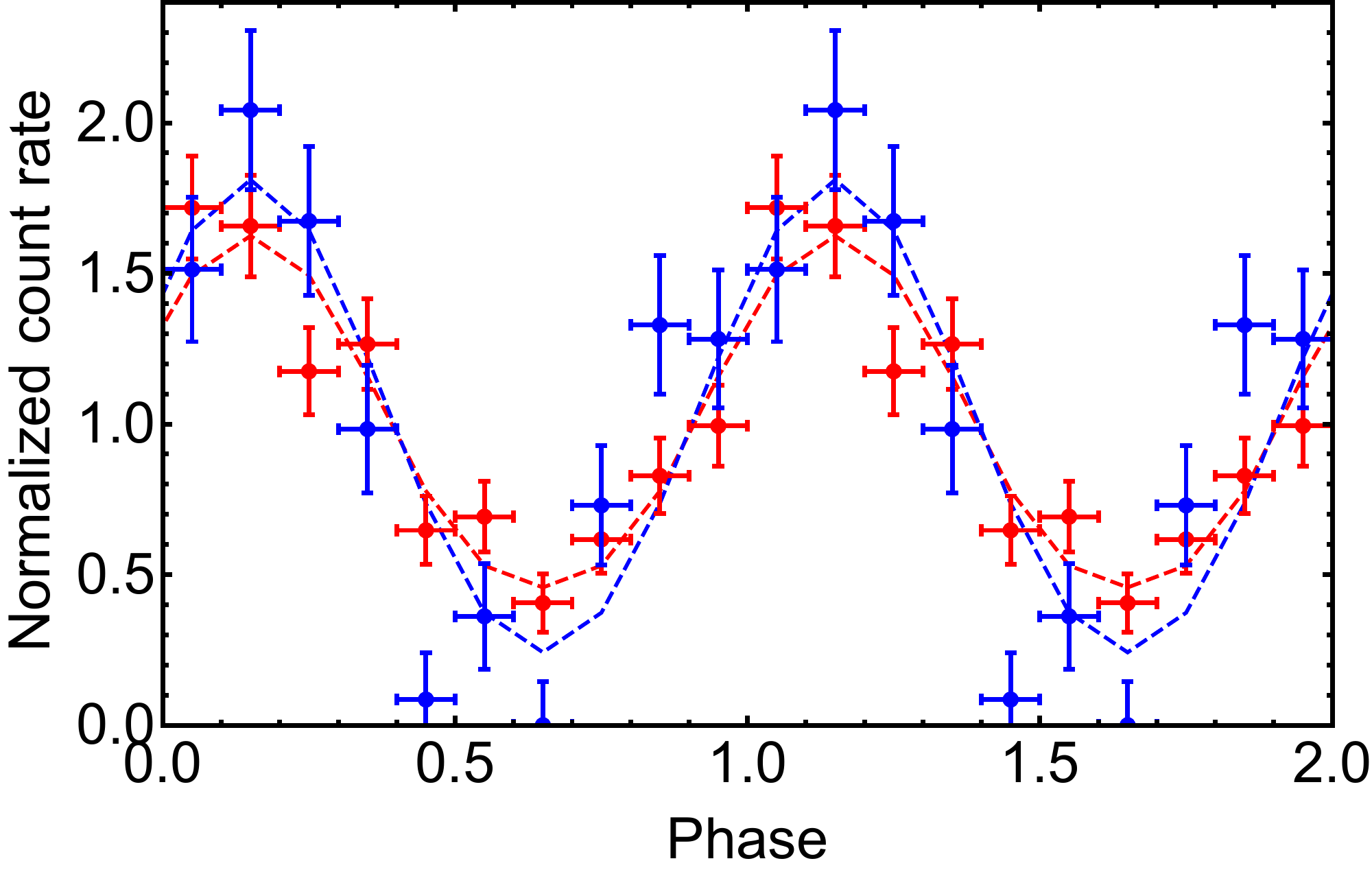}}
  \caption{Background subtracted pulse profile of SGR~0418 in the
    energy range 0.3--1.2\keV\ (red) and 1.2--10.0\keV\ (blue).  The
    dashed line corresponds to the best synthetic pulse profiles
    obtained from the grid search described in
    Section~\ref{sec:iterative} for each of the two energy bands.}
  \label{fig:BestFitPulseProfilesTwoBands}
\end{figure}

A consistency check was then performed by using the values of the
best-fit angles $\left(124\deg,4\deg\right)$ and temperature
distribution parameters ($n=7500$ and $\epsilon=0.10$) in a
phase-averaged spectral fit.  These parameters lead to a best-fit
temperature $\tpinf=0.34\ud{0.03}{0.03}\keV$ and a hydrogen column
density $\nhtt=0.30\ud{0.06}{0.04}$, for a corresponding distance of
$0.5\kpc$.  The pole temperature and the hydrogen column density
remain consistent with the best-fit $\tpole$ obtained when fixing
$\left(\psi,\xi\right)=\left(90\deg,90\deg\right)$, as done at the
beginning of this iterative analysis.  In addition, for the best-fit
temperature distribution found above ($n=7500$ and $\epsilon=0.10$),
the viewing geometry appears to have an effect on the distance
(i.e. flux normalization). For
$\left(\psi,\xi\right)=\left(124\deg,4\deg\right)$,
$d=0.5\ud{0.1}{0.1}\kpc$ and for
$\left(\psi,\xi\right)=\left(90\deg,20\deg\right)$,
$d=2.7\ud{1.0}{0.9}\kpc$ (these uncertainties are essentially flux
uncertainties).  While this was not suspected initially, this
dependence of the distance on the viewing geometry is explained by the
fact that a highly contrasted spot, more or less-often visible during
a complete phase depending on the viewing angles, will correspondingly
generate a larger or smaller flux in the phase averaged spectra.

\begin{figure}
  \centering
  \makebox[0cm]{\includegraphics[width=8cm]{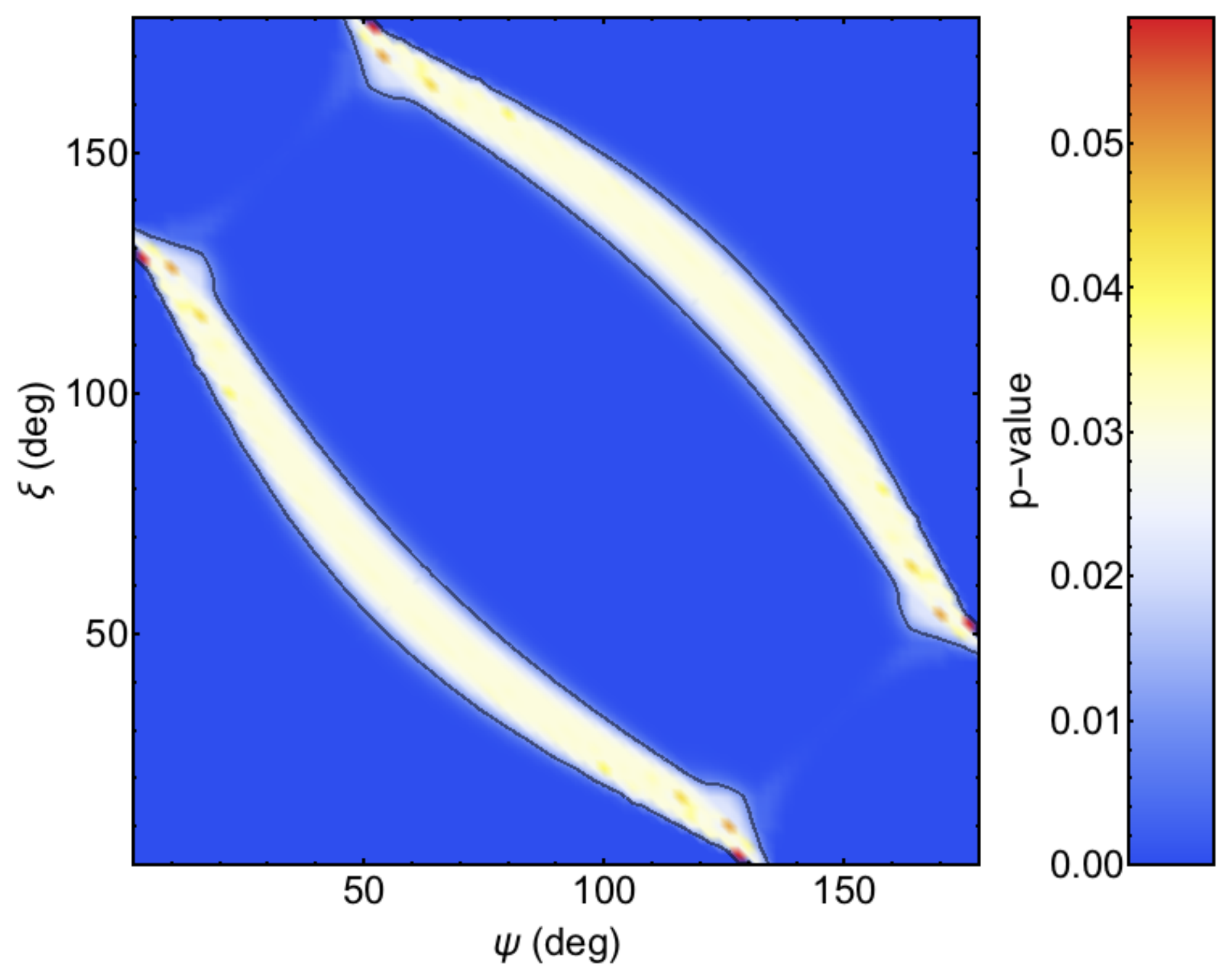}}
  \caption{ Same as in Figure~\ref{fig:beaming}, but in the case of an
    isotropic local emission.}
  \label{fig:nobeaming}
\end{figure}

The grid search described above was then repeated with the pulse
profiles computed in the two energy bands 0.3--1.2\keV\ and
1.2--10.0\keV.  The resulting best fits pulse profiles are shown in
Figure~\ref{fig:BestFitPulseProfilesTwoBands}, and the 2D projections
of maximum p-values are displayed in Figure~\ref{fig:twobands}.  The
analysis of the pulse profiles in the two bands shows that the higher
energy bands constrain the angles a little more than the low energy
band and this is simply due to the larger pulse fraction (consistent
with 1.0) observed in the 1.2--10.0\keV\ band compared to the
0.3--1.2\keV\ band.  But overall, the parameter space is mostly
insensitive to the energy band chosen, given the low signal-to-noise
available in these obervations.

Last, with the purpose of quantifying the effect that the uncertain
atmospheric beaming has on our results, we repeat the analysis
assuming local isotropic emission.  This results in a narrower allowed
parameter space for the angles $\psi$ or $\xi$. For example, we find
that the angles $\psi$ or $\xi$ must lie in the range
$104\deg\simlt\psi+\xi\simlt145\deg$ and $215\deg\simlt\psi+\xi\simlt
256\deg$.  Moreover, the contours are less curved in $\psi$--$\xi$
space (Figure~\ref{fig:nobeaming}).  These results can be understood
since the effect of beaming is that of creating a larger modulation
for the same parameters. Hence, to reproduce the same level of
observed modulation with a local isotropic emission beaming, smaller
viewing angles are not allowed, restricting the angle geometry to
larger values.

\begin{figure}
  \centering
  \makebox[0cm]{\includegraphics[width=8cm]{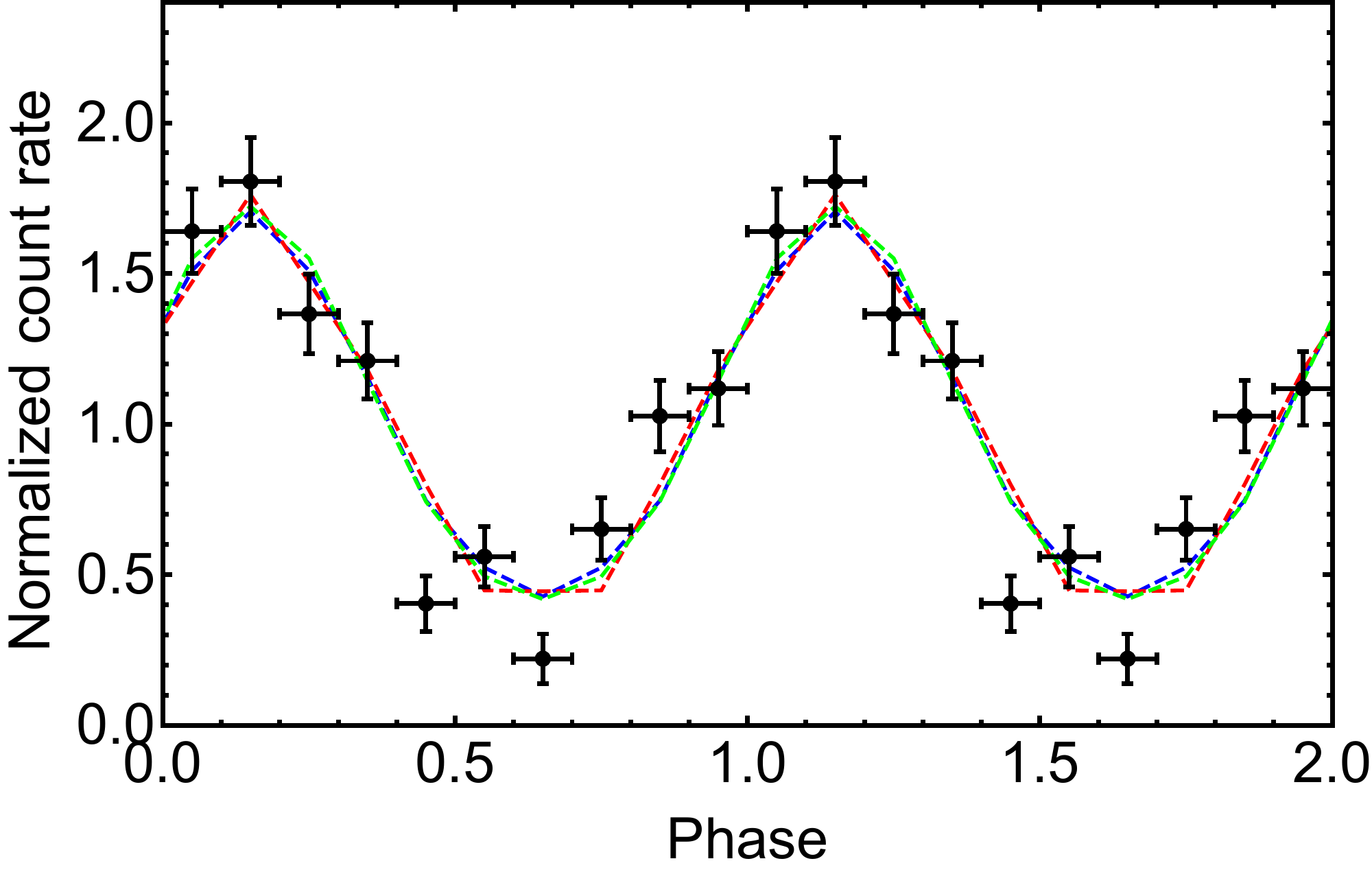}}
  \caption{Background subtracted pulse profile of SGR~0418, together
    with the best fit synthetic pulse profiles obtained from the grid
    search described in Sections~\ref{sec:iterative} and
    \ref{sec:twospots} (dashed blue: single spot with pencil-beamed
    surface emission, dashed red: single spot with isotropic surface
    emission, dashed green: two spots with pencil-beamed surface
    emission).  They are computed in the 0.3--10\keV\ band.}
  \label{fig:BestFitPulseProfiles}
\end{figure}

The best-fit pulse profiles in the locally isotropic
[\Chisq{1.9}{8}{0.06}] and pencil-beamed surface emission are compared
in Figure~\ref{fig:BestFitPulseProfiles}.  As it can be seen, the
profiles are almost indistinguishable. This result stems from what
just discussed above: when the local emission is beamed, the fit
parameters adjust so that smaller viewing angles/temperature contrast
are needed to achieve the same observed level of modulation.
Therefore, a comparison between Figure~\ref{fig:beaming} and
\ref{fig:nobeaming} yields a quantitative estimate of the amount by
which the uncertain angular distribution of the local emission affects
the inferred system parameters.

\subsection{Simulateneous fitting of spectra and pulse profiles using a Markov-Chain Monte-Carlo approach}
\label{sec:mcmc}

As explained in Section~\ref{sec:iterative}, the source spectrum is
more sensitive to some parameters of the model (e.g.: \tpole\ or the
spot size charaterized by $n$ in Equation~\ref{eq:Tprofile}) while the
pulse profile is more sensitive to variations in the temperature
contrast at the surface or in the system geometry ($\psi$ or $\xi$).
Nonetheless, all parameters play a role in modelling the surface
emission and its phase- and energy- dependence, and
Section~\ref{sec:iterative} showed that it was somewhat difficult to
predict what variations of a given parameter can have on the modelled
pulse profile or spectrum.  Iterating between pulse profile and
spectral fits proved to be impractical due to the number of
parameters.  Furthermore, a simultaneous analysis allows us to have a
global understanding on the roles of parameters in the model, and to
better understand the existing degeneracies between parameters.  For
these reasons, a Markov-Chain Monte Carlo (MCMC) approach was then
employed to sample the parameter space while simultaneously fitting
the spectra and pulse profiles.  In addition, the method allows to
include known priors on some of the parameters.

\begin{figure*}
  \centering
  \makebox[0cm]{\includegraphics[width=18cm]{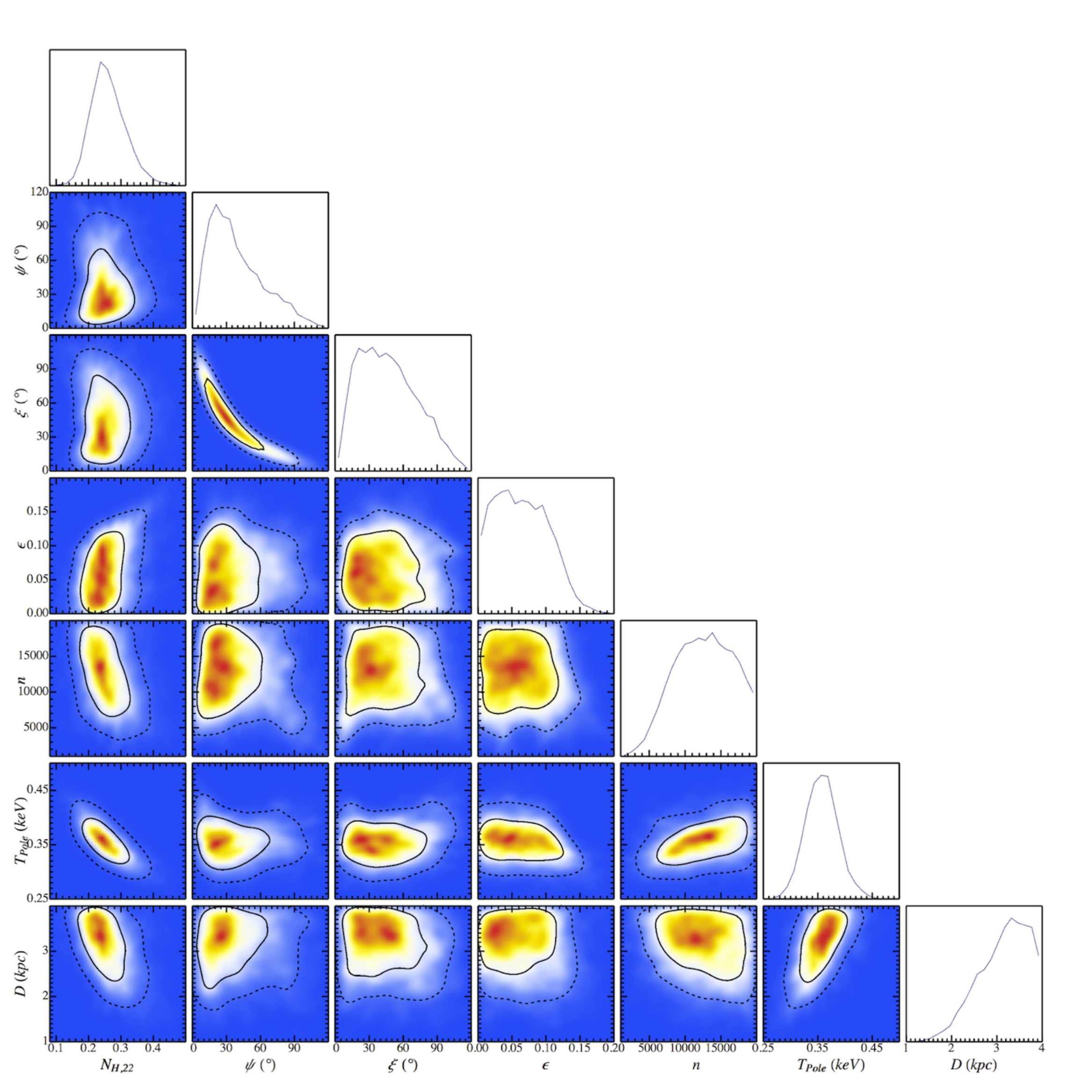}}
  \caption{2-dimensional posterior distributions for each pair of
    parameters and 1-dimensional posterior distributions for each
    parameter resulting from the simultaneous fits of the pulse
    profiles and spectra via the MCMC approach described in
    Section~\ref{sec:mcmc}.  The color scales of the 2-dimensional
    posterior distributions represent the density of points (i.e.,
    blue means zero-density), compared to other figures of this
    article where the color scales represent the p-values.  The solid
    and dashed contours are enclosing 68\% and 95\% of the accepted
    points. This figure was created with the Mathematica package
    LevelSchemes \citep{caprio05}.}
  \label{fig:mcmc}
\end{figure*}

The implementation of the MCMC algorithm used in this work, the
``Stretch-Move'' algorithm \citep[also called ``Affine Invariant
  Ensemble Sample'',][]{goodman10}, was previously employed, described
and tested extensively in other works
\citep[e.g.,][]{wang13,guillot13,guillot14}.  It made use of the
\emph{PyXSPEC} package, the python version of \emph{XSPEC}
\citep{arnaud96}, for the comparison of the synthetic spectra to the
phase-average spectra.  In this algorithm, multiple chains (called
``walkers'') are sampling the parameter space and each set of proposed
parameters is accepted or rejected based on the likelihood of the
model with the proposed sets of parameters.  The next proposed step of
each chain is then chosen from an affine invariant distribution along
a line between the current point of the chain and the current point of
a randomly chosen chain\footnote{The description of the algorithm and
  its performance are described in \citet{goodman10}.}.

In this approach, we used a set of seven parameters sampled by the
MCMC:
\begin{itemize}
\item {the two angles of the system geometry $\psi$ and $\xi$
  (constrained to $\psi+\xi<180\deg$, as justified by the symmetry, in
  Eq.~(\ref{eq:alpha}); this is to reduce the size of the parameter
  space explored, and the full range can be recovered by symmetry).}
\item {the surface temperature contrast $\epsilon$ (Eq.~\ref{eq:Tprofile}),}
\item {the parameter $n$ characterizing the spot size
  (Eq.~\ref{eq:Tprofile}),}
\item {the column density of hydrogen \nh,}
\item {the pole temperature \tpole},
\item {the distance $d$, for which we choose a flat prior between 0.5
  and 4\kpc\,(see discussion in Section~\ref{sec:iterative}).}
\end{itemize}
The neutron star radius, which controls the star's compactness and
therefore the amount of light-bending, could also be sampled in this
approach to potentially obtain some constraints on its value.
However, it was held fixed here due to the relatively low signal-to-noise
ratio of the data. Higher signal-to-noise data could potentially place
constraints on the radius, although there exists degeneracies with
the angle geometry that would not be fully broken (see e.g.,
\citealt{perna08,bernardini11}).

\begin{figure*}
  \centering
  {\includegraphics[width=8cm]{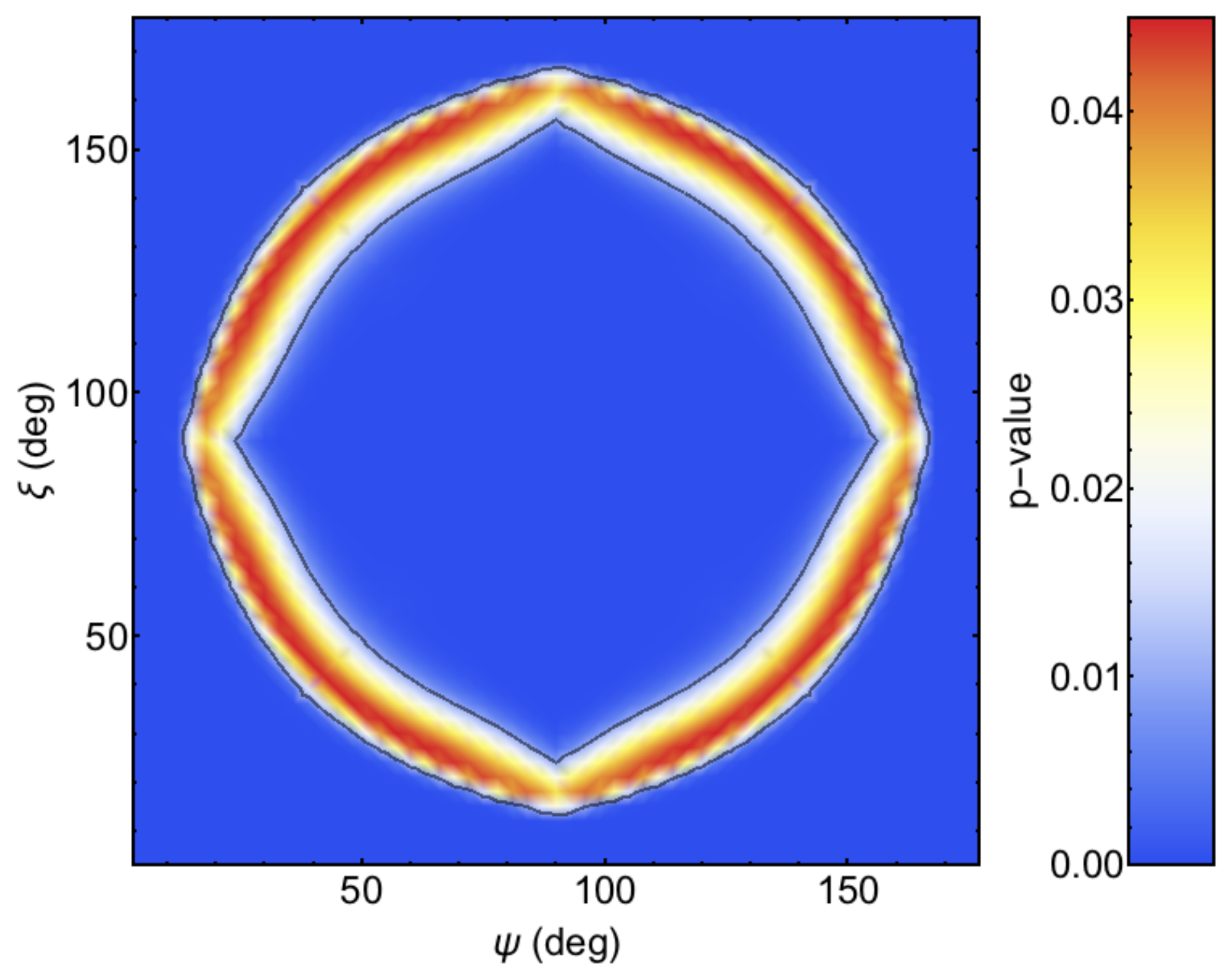}}
  {\includegraphics[width=8cm]{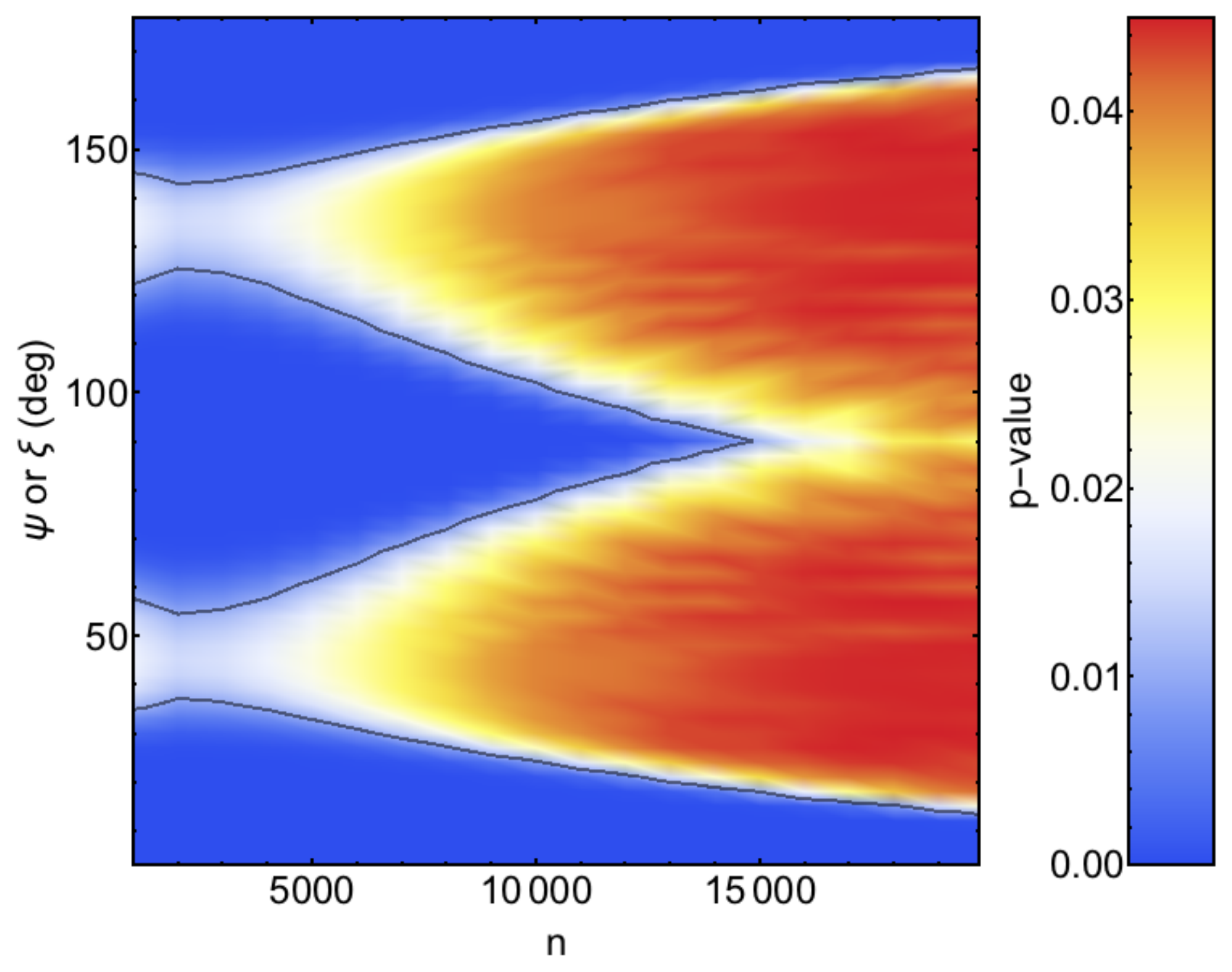}}
  \caption{ ({\it left}) Similar to Figure~\ref{fig:beaming} but for a
    symmetry temperature distribution representing two antipodal
    spots. The solid line represents the probability $p=0.01$.  ({\it
      right}) Projected map of maximum p-values in a $n$--$\psi$ or
    $n$--$\xi$ space. }

  \label{fig:2spots}
\end{figure*}

In this MCMC simulation, 50 walkers were run simultaneously for a
total of 15000 steps each.  After removing 5000 steps of burn-in, an
average acceptance rate of 15\% was obtained.  We also checked for
convergence by visually inspecting the trace of each parameter.  The
results of this simultaneous analysis of the pulse profiles and
spectra of SGR~0418 with our model are shown in Figure~\ref{fig:mcmc},
as the 1-dimensional and 2-dimensional posterior distributions of all
parameters.

This simultaneous analysis confirms what has been observed in the
iterative method of Section~\ref{sec:iterative}.  In particular, the
constraint on the geometry of the system (see the $\psi-\xi$ plot of
Figure~\ref{fig:mcmc}) is well reproduced, albeit with a slightly
different shape that results from the coupled constraints from the
spectra (as noted in Section~\ref{sec:iterative}, the distance and the
angle geometry are related).  The constraints on the two angles are:
$65\deg\simlt\psi+\xi\simlt125\deg$ or $235\deg\simlt\psi+\xi\simlt
295\deg$.

Furthermore, because a conservative prior on the distance was
included, the size of the spot (parametrized via $n$) is not
particularly well constrained.  We find that the spot size is in the
range 1.2\deg--3.8\deg, equivalent to $0.2-0.7\km$ for the 10-km NS
considered here.  The spot size distribution peaks at $0.35\km$.  We
note that allowing for larger distances would therefore include larger
spot sizes (i.e, smaller values of $n$) in the posterior
distributions.  Since this is driven by the spectra normalization, it
would not affect significantly the posterior distributions of
\tpole\ or \nh.  However, a consequence would be a broadening of the
``banana''-shape contours of the 2D-distribution of $\psi$--$\xi$,
since a larger spot would reproduce the pulse profiles for larger
values of the angles $\psi$ and $\xi$.  Note also that because of the
symmetry arising from the system geometry, the angles in the MCMC run
were constrained to $\psi+\xi<180\deg$, and therefore, the $\psi-\xi$
plot only shows one of the ``banana'' shape contours.

The temperature contrast $\epsilon$ is constrained to similar values
as those obtained in the iterative analysis, i.e. $\epsilon\simlt
0.15$.  However, it is important to note that while allowed here,
$\epsilon \rightarrow 0.0$ represents an unphysical description of the
temperature distribution at the surface of the NS since the cold part
cannot have a zero temperature.  Furthermore, given the posterior 2D
distributions of Figure~\ref{fig:mcmc}, one can readily see that
excluding the lower range of $\epsilon$, say 0.00--0.05, would not
significantly affect the other parameters due to the mild correlation
between $\epsilon$ in this range and the other parameters.

The posterior distribution of the pole temperature is
$\tpinf=0.36\pm{0.05}\keV$ (90\% c.l.) while the hydrogen column
density is $\nhtt=0.25\ud{0.12}{0.08}$ (90\% c.l.).  Note that we
initially observed the bimodal distribution of \nh\ and therefore
$\epsilon$ (leading to the secondary \chisq\ minimum in the spectral
fit described in Section~\ref{sec:iterative}).  We then applied a
prior on \nh\ to exclude values $\nhtt>0.8$ (see
Section~\ref{sec:iterative}).  The distance is slightly more
constrained than the initially prior range allowed.  We find a
distribution between 1.5\kpc\ and 4\kpc, skewed toward larger values.
Nonetheless, the 2\kpc\ distance suggested if SGR~0418 indeed resides
in the Perseus arm cannot firmly excluded.

Finally, the best-fit is obtained for the following set of parameters:
$\nhtt=0.24$, $\psi=96.8\deg$, $\xi=7.9\deg$, $\epsilon=0.08$,
$n=16961$, $\tpinf=0.38\keV$ and $d=2.27\kpc$, corresponding to a fit
statistic \Chisq{58.5}{38}{0.02}.  However, it is crucial to keep in
mind that the full posterior distributions are the true representation
of the acceptable parameter space given the model chosen and the data.

Overall, this technique is superior to the iterative process of
Section~\ref{sec:iterative} since it includes all the effects of
parameter variations on both the pulse profiles and spectra, which
proved too difficult to perform iteratively.  Nonetheless, the
iterative analysis performed initially provided a validation of the
simultaneous analysis via MCMC.

\subsection{Can two antipodal spots be responsible for the observed pulse profile of SGR~0418?}
\label{sec:twospots}

We present here an analysis similar to that of
Section~\ref{sec:iterative} but with a symmetric temperature
distribution at the surface of the magnetar
(Equation~\ref{eq:Tprofile2}), i.e., with two antipodal spots.  As can
be seen in Figure~\ref{fig:2spots} (left panel), the two-spots
temperature distribution is able to accomodate the observed pulse
profile for a narrow range of angles.  One can note the additional
symmetry (compared to the single spot case) since the two spots are
identical.  The two angles $\psi$ and $\xi$ are constrained to be
$\simgt 15\deg$ and $\simlt 165\deg$, although the value of one angle
strictly constrains the value of the other (and the two angles are
symmetric with respect to exchange).  As with the single spot case,
$\epsilon\simlt 0.15$, i.e. $\epsilon$ is not constrained more than
with the spectral analysis.

However, we observed a strong correlation between the spots size and
the angle geometry. Specifically, the larger the two spots, the more
restricted the angles (see Figure~\ref{fig:2spots}, right panel).
This can easily be explained by the fact that two large spots (small
$n$) can only reproduce the single-peaked pulse profile observed for
angles $\psi$ and $\xi$ with values $\sim 45\deg$ or $\sim 135\deg$.
However, as the spots get smaller, they can accommodate the observed
pulse fraction and profiles for wider ranges of angles.  This results
from the combined effect of beamed emission and light bending which
allow the observer to see one spot appearing from behind the neutron
star while the second spot is still visible.

In this case of symmetric hot spots, the best fit to the pulsed
profile is obtained for the parameters
$\left(\psi,\xi\right)=\left(57\deg,30\deg\right)$, and
$\epsilon=0.05$ and $n=18000$.  As in Section~\ref{sec:iterative}, we
perform a consistency check using these parameters for a spectral fit.
We obtain the best fit $\tpinf=0.35\pm0.05\keV$,
$\nhtt=0.27\ud{0.10}{0.08}$, a corresponding distance
$d=1.5\ud{0.7}{0.5}\kpc$, and a fit statistic
\Chisq{1.26}{34}{0.15}. Figure~\ref{fig:BestFitPulseProfiles} shows
the best-fit pulse profile obtained with the symmetric temperature
distribution at the surface of SGR~0418, alongside with the best-fit
synthetic pulse profiles for the one-spot models (beaming and
isotropic local emissions).

Overall, we find that two identical small spots with a beamed local
emission can produce high PFs and single peaked pulse profiles.
Therefore, observed pulse profiles can be fitted equally well with a
symmetric temperature distribution (two spots) as with asymmetric
models (single spots), hence making the two situations equivalent.
This result is similar to that derived for other sources
\citep{shabaltas12,storch14}.

\section{Interpretation of our findings within the context of heating models}
\label{sec:magneto}

The strongest constraint of our modelling is the presence of a high
contrast in the temperature distribution on the surface of the star.
The hottest point on the star, \tpinf, is inferred to be at a value of
about 0.35\keV. The temperature then declines to an antipoidal minimum
which has been constrained to be at least a factor of $\sim 6$ smaller
than the maximum, but which can be much smaller (note that emission
from the coolest part of the star is not detectable in our
observations). The precise rate of decline between the maximum and
minimum values is not well determined by the current data. However,
for the range of acceptable temperature profiles and distances, there
is a region around the pole with an angular size of a few degrees for
which the temperature hovers above $\sim 0.3\keV$.

The inferred high temperature of the region dominating the emission is
difficult to reconcile with the standard cooling model of NSs, even
considering magneto-thermal evolutionary models with extra heating by
magnetic field decay.  The estimated age of the star, consistent with
its luminosity and spin evolution, is about half a million years
\citep{rea13}. At that age, the temperatures of the hottest regions in
a normal pulsar are expected to be below 0.1\keV, with some dependence
on the equation of state, NS mass, superfluid gap, envelope model,
etc. This result also holds for the available models with very strong
initial poloidal fields $B_p \sim 10^{15}\G$ \citep{vigano13}, or weak
dipolar field with extremely strong initial toroidal fields, $B_t \sim
10^{16}\G$ \citep{geppert14}, which is qualitatively more compatible
with the timing properties and the outburst activity of SGR 0418. With
some tuning of the microphysics parameters, the expected temperature
of the hot spot at 0.5 Myr may be increased up to a 50\% (say
0.15\keV).  The mismatch between theory and observation could be
somewhat compensated by atmospheric effects, which may give a color
correction by about a factor $\sim 2$ (e.g. \citealt{lloyd03}), and
hence could possibly account for inferred temperatures up to
0.3\keV. However, the innermost region of the spot, of about a degree
in size and temperature above 0.3\keV\, cannot be explained by
residual cooling and internal heating in a 0.5 Myr old NS.  We note
that rather high temperatures accompanied by small inferred radii of
the emitting regions are a common issue to most magnetars. In
relatively young objects with strong toroidal fields and high $N_{\rm
  H}$, magnetothermal simulations show that km-sized hot spots can in
fact be produced \citep{perna13}, but this is problematic in the case
of SGR~0418 since it is much older than the rest of the sources, and
the size of the hottest region is especially small.

There are different possibilities to explain the anomalously high
temperature. The first one is simply that the object has not reached
its quiescence state yet; however, since the observations have showed
a steady flux for more than a year, we believe that quiescence has
indeed been reached.  Therefore, unless something is missing from our
theoretical understanding of magnetized NS cooling \citep{pons09} or
from the physics of the NS crust and envelope, we conclude that the
hotspot must be maintained by an external heating source, attributed
to energetic particles carried by magnetospheric currents and falling
into the polar cap. The bombardment of these particles could account
for the large, persistent temperature with the caveat of how to
maintain stable current systems on a timescale of years. At present
only analytical estimates of the bombardment energetics are available,
without any detailed numerical simulation. Thus, deeper investigations
are needed to conclude in favour of one or another option.

\section{Conclusions}
\label{sec:ccl}

We have modelled the post-outburst surface emission of SGR~0418+5729.
The low-luminosity emission, observed with \xmmlong\ in 2012 and 2013
is consistent with being constant on this time-scale, hence indicating
that the source has reached quiescence. Using a general-relativistic,
phase-dependent thermal spectral model, we have fit the spectrum and
pulse profile of SGR~0418 to constrain the geometry of the system as
well as the temperature distribution profile on the surface of the
NS. 

This analysis was performed using two independent analyses: one
approach which requires iteration between spectral analysis and pulse
profile fitting to constrain the parameters of the model; a second
method using a Markov-Chain Monte Carlo approch that simultanesously
fitted the pulse profile and spectra of SGR~0418.  The two methods led
to consistent results.

We have found that SGR~0418 has a high temperature contrast on the
surface, with differences between the maximum and the minimum
temperature of a factor of at least $\sim6$.  Despite the
single-peaked pulse profile, the possibility of a symmetric
temperature distribution at the surface of SGR~0418 (i.e., two
antipodal spots) cannot however be excluded.  The small size of the
spots, combined with radiation beaming and the absorption of soft
X-rays, allow for a high PF single-peak pulse profile observed for
SGR~0418. Significant constraints were also placed on the
viewing/emission geometry.  While each angle can take any value
between 0 and 180\deg, we constrained $\psi$ and $\xi$ in a correlated
manner such that $65\deg\simlt\psi+\xi\simlt125\deg$ or
$235\deg\simlt\psi+\xi\simlt 295\deg$ (with a mild dependence on the
radiation beaming; for isotropic emission the requirement would be
more constraining: $104\deg\simlt \psi+\xi\simlt 145\deg$ or
$215\deg\simlt \psi+\xi\simlt 256\deg$).

The inferred value of the NS surface temperature, exceeding
0.3\keV\ in a region of about a few degrees in size, is very difficult
to explain in a source of about 0.5 Myr with standard cooling models,
even accounting for possible color corrections due to atmospheric
processing of the emitted radiation, and varying the micro and macro
physics parameters of the currently available cooling models. While a
contribution from internal heating in the cooler region of the star
cannot be excluded, we believe that the dominant, small hot spot must
be maintained by the bombardament of energetic particles carried by
magnetospheric currents.

\vspace{0.5cm}

{\bf Acknowledgements.} We thank the anonymous referee for a careful
reading of our manuscript and insightful comments.  SG acknowledges
the hospitality of Joint Institute for Laboratory Astrophysics (JILA,
at the University of Colorado) where this work was initiated. SG is
funded by the FONDECYT postdoctoral grant 3150428.  During the
preparation of this work, SG was partially funded at McGill University
by NSERC via the Vanier Canada Graduate Scholarship program, and by
the Fonds de Recherche du Qu\'{e}bec - Nature et Technologies.  RP
acknowledges support by NSF grant No. AST 1414246 and
Chandra-Smithsonian Awards No. GO3-14060A and No. AR5-16005X.  NR
acknowledges support via an NWO Vidi Award.  NR and DV are supported
by grants AYA2012-39303 and SGR2009-811. Partial support comes from
NewCompStar, COST Action MP1304.  JP is supported by the grant
AYA2013-42184-P.

\bibliographystyle{apj_8} 
\bibliography{biblio}

\end{document}